\documentclass[aps,twocolumn,pre,eqsecnum]{revtex4} 
\usepackage{graphpap}
\pdfoutput=1
\usepackage[dvips]{graphicx}
\usepackage[dvips]{graphics}
\usepackage{color}
\usepackage[normalem]{ulem}
\usepackage {ulem}
\usepackage{amsmath}

\usepackage{soul}

\begin{document}

\title{Line Shape of the Raman 2D Peak of Graphene in Van Der Waals Heterostructures}

\author{C. Neumann$^{1,2}$, L. Banszerus$^1$, M. Schmitz$^1$, S. Reichardt$^{1,3}$, J. Sonntag$^{1,2}$, K.~Watanabe$^4$, T.~Taniguchi$^4$, B. Beschoten$^1$,  and C. Stampfer$^{1,2}$}
\affiliation{
$^1$\,JARA-FIT and 2nd Institute of Physics, RWTH Aachen University, 52074 Aachen, Germany\\
$^2$\,Peter Gr\"unberg Institute (PGI-9), Forschungszentrum J\"ulich, 52425 J\"ulich, Germany\\
$^3$\,Physics and Materials Science Research Unit, Universit\'e du Luxembourg, 1511 Luxembourg, Luxembourg\\
$^4$\,National Institute for Materials Science,1-1 Namiki, Tsukuba, 305-0044, Japan
}
\date{ \today}

\begin{abstract}
The Raman 2D line of graphene is widely used for device characterization and during device fabrication as it contains valuable information on e.g. the direction and magnitude of mechanical strain and doping. Here we present systematic asymmetries in the 2D line shape of exfoliated graphene and graphene grown by chemical vapor deposition. Both graphene crystals are fully encapsulated in van der Waals heterostructures, where hexagonal boron nitride and tungsten diselenide are used as substrate materials. In both material stacks, we find very low doping values and extremely homogeneous strain distributions in the graphene crystal, which is a hall mark of the outstanding electronic quality of these samples. By fitting double Lorentzian functions to the spectra to account for the contributions of inner and outer processes to the 2D peak, we find that the splitting of the sub-peaks, $6.6 \pm 0.5$~cm$^{-1}$ (hBN-Gr-WSe$_2$) and $8.9 \pm 1.0$~cm$^{-1}$ (hBN-Gr-hBN), is significantly lower than the values reported in previous studies on suspended graphene.
\end{abstract}

\maketitle   % please do not remove

\section{Introduction}
Raman spectroscopy has emerged as a key tool for studying low dimensional carbon systems such as carbon nanotubes and graphene \cite{ferrari2013}.
In particular, the Raman 2D peak of graphene (Gr) has been subject to numerous experimental \cite{thomsen2000,ferrari2006,maultzsch2004,graf2007} and theoretical studies \cite{basko2008,venezuela2011,narula2012} over the past years. A precise understanding of the intrinsic properties of the Raman peak is highly desirable, since the 2D peak is frequently used for sample characterization and analysis in graphene research, synthesis and device fabrication \cite{ferrari2013,neumann2015b,banszerus2015}. Most notably, detailed information on the number of graphene layers \cite{ferrari2006,graf2007}, the amount and nature of strain \cite{lee2012,mohiuddin2009,huang2010,frank2011}, and the doping in a graphene sample can be extracted from the 2D peak \cite{lee2012,froehlicher2015,das2008}. In most experimental studies the Raman 2D peak of single-layer graphene is modeled by a single Lorentzian function and the frequency (i.e. position) and broadening of the Lorentzian are utilized for data interpretation \cite{ferrari2006,graf2007}. However, studies on clean suspended graphene samples have revealed intrinsic asymmetries in the line shape, revealing that the processes contributing to the Raman 2D peak cannot be fully captured by a single Lorentzian \cite{berciaud2008,luo2012,berciaud2013} since the contributing processes are not indistinguishable.
Here we report reproducible asymmetries in the line shape of high-quality graphene encapsulated in van der Waals heterostructures, where hexagonal boron nitride (green data in this manuscript) and tungsten diselenide (blue data in this manuscript) are used as substrate materials. We find that the 2D peaks obtained in clean areas in such van der Waals heterostructures are even narrower than the narrowest peaks reported for suspended graphene. This is a clear indication that local strain variations within the laser spot are highly suppressed in these material stacks \cite{neumann2015b,woods2014,eckmann2013}.
Following approaches to describe the 2D peak by a superposition of two Lorentzian sub-peaks (related to inner and outer scattering processes as described later in the manuscript) \cite{venezuela2011,luo2012,berciaud2013}, we analyse the intensity ratio and frequency difference of the two sub-peaks.
We find that the Raman 2D line shape in ultra clean van der Waals heterostructures resembles the shape found for clean, suspended graphene with the most striking difference being the smaller full-width at half-maximum (FWHM) and a smaller frequency difference of the of the sub-peaks. 

\section{Methods}
The Raman spectra are obtained with a confocal, scanning Raman setup, which is equipped with a x-y-z DC piezo stage. For the measurements, two different objectives, a 50x objective with NA=0.7 and a 100x long working distance objective with NA=0.75 are employed. For excitation a 532~nm (2.33~eV) laser is used. For detection the light is guided via a single mode optical fiber to a CCD array spectrometer with a grating of 1200 lines/mm.

\begin{figure}[hbt]
\centering
\includegraphics[draft=false,keepaspectratio=true,clip,width=1\linewidth]{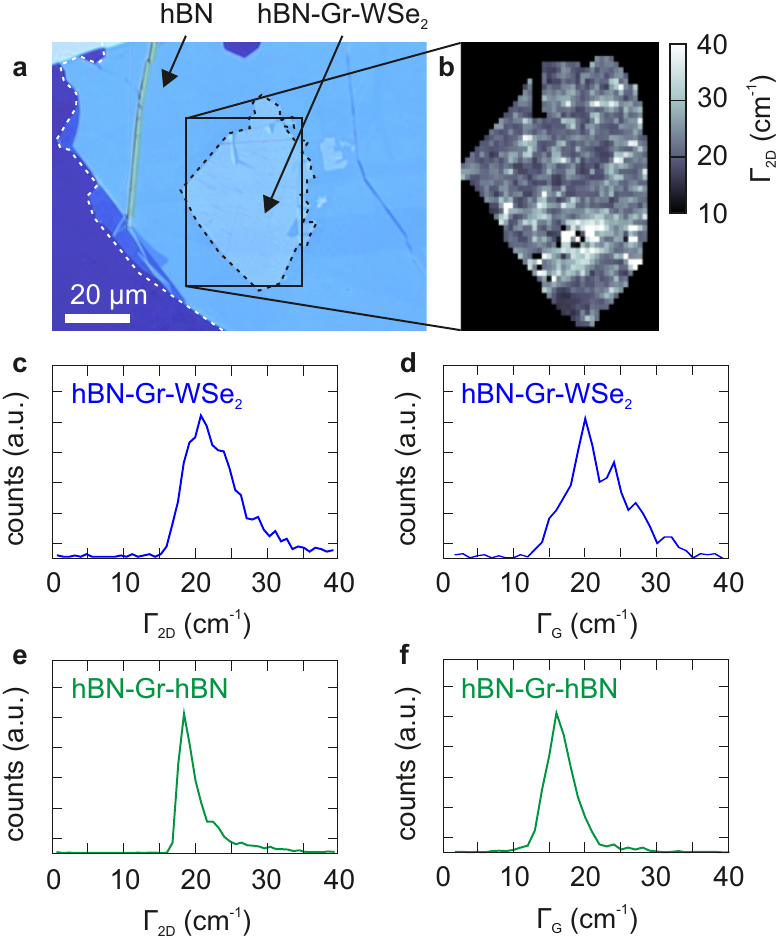}
\caption[Raman spectroscopy characterization of hBN-Gr-WSe$_2$ and hBN-Gr-hBN]{\label{fig:shapesample}
(a) Optical microscopy image of a hBN-Gr-WSe$_2$ sample. The white, dashed line highlights the contours of the hBN crystal. The black, dashed lines highlights the area where the CVD Gr is encapsulated between WSe$_2$ and hBN. 
(b) Scanning Raman map displaying the FWHM of the 2D line as obtained from single Lorentzian fits to every spectrum.
(c) Histogram of $\Gamma_{2D}$ obtained from the Raman map shown in (b).
(d) Histogram of $\Gamma_{G}$ as extracted from single Lorentzian fits to the G peak from the Raman map used in (b,c).
(e) Histogram of $\Gamma_{2D}$ obtained from a Raman map on an hBN-Gr-hBN heterostructure.
(f) Histogram of $\Gamma_{G}$ obtained from the Raman map used in (e). All measurements shown in this figure are based on a laser energy of 2.33~eV.}
\end{figure}
\section{Measurements and Discussion}
In Figure \ref{fig:shapesample}a an hBN-Gr-WSe$_2$ (Figure~\ref{fig:shapesample}b) located on an Si$^{++}$/SiO$_2$ chip are shown. The sample is fabricated from chemical vapor deposited (CVD) graphene grown on copper foil. The graphene is released from the copper foil and picked up with the WSe$_2$ crystal following the procedure described by Banszerus \textit{et al}. \cite{banszerus2015}. In the following, the material stack is placed onto an hBN crystal which is located on an Si$^{++}$/SiO$_2$ chip. This particular method of producing van der Waals heterostructures from CVD graphene leads to extremely clean graphene samples with very low doping and extremely high electronic quality, indistinguishable from state-of-the-art exfoliated samples \cite{banszerus2015,neumann2015c,banszerus2016}. We compare our results obtained on the hBN-Gr-WSe$_2$ sample, with measurements taken on an hBN-Gr-hBN stack that is fabricated using mechanically exfoliated graphene by a wet chemistry-free transfer process \cite{wang2013,engels2014b}, resulting in similarly clean graphene samples with outstanding electronic quality \cite{wang2013,engels2014b,neumann2016}. In general, both hBN and WSe$_{2}$ are known to be very suitable substrates for graphene leading to very flat graphene layers with high charge carrier mobilities \cite{xue2011,dean2010,kretinin2014}.
In Figure \ref{fig:shapesample}b a scanning Raman map displaying the FWHM of the 2D peak, $\Gamma_{2D}$, obtained from single Lorentzian fits for the hBN-Gr-WSe$_2$ sample is shown. A corresponding histogram of $\Gamma_{2D}$, obtained from this Raman map, is presented in Figure \ref{fig:shapesample}c. The lowest line widths that we can find on the sample are around 15-16 cm$^{-1}$, which proves the very uniform strain distribution within the laser spot \cite{neumann2015b,neumann2015}. It is noteworthy that the line width of the 2D line that we obtain in our hBN-Gr-WSe$_2$ structures is even smaller than the smallest values reported for suspended graphene \cite{berciaud2013}, meaning that averaging effects over different local strain conditions are highly suppressed in our van der Waals heterostructures. For studying the intrinsic line shape of the 2D peak, this high strain uniformity is a fundamental prerequisite as strain variations within the laser spot introduce significant broadening to the 2D peak that would mask the intrinsic properties. In Figure \ref{fig:shapesample}d, we show a histogram of the FWHM of the G peak, $\Gamma_{G}$, obtained from single Lorentzian fits to the G peaks of a Raman map of the entire sample. The high values of $\Gamma_{G}$ indicate the generally very low doping of the graphene layer in this heterostructure \cite{yan2007,pisana2007,froehlicher2015}. 
The strain uniformity and doping conditions in the investigated hBN-Gr-hBN sample are very similar as evident from the two histograms of $\Gamma_{2D}$ (Figure \ref{fig:shapesample}e) and $\Gamma_{G}$ (Figure \ref{fig:shapesample}f) obtained from a scanning Raman map of this sample.

\begin{figure*}[hbt]
\centering
\includegraphics[draft=false,keepaspectratio=true,clip,width=1\linewidth]{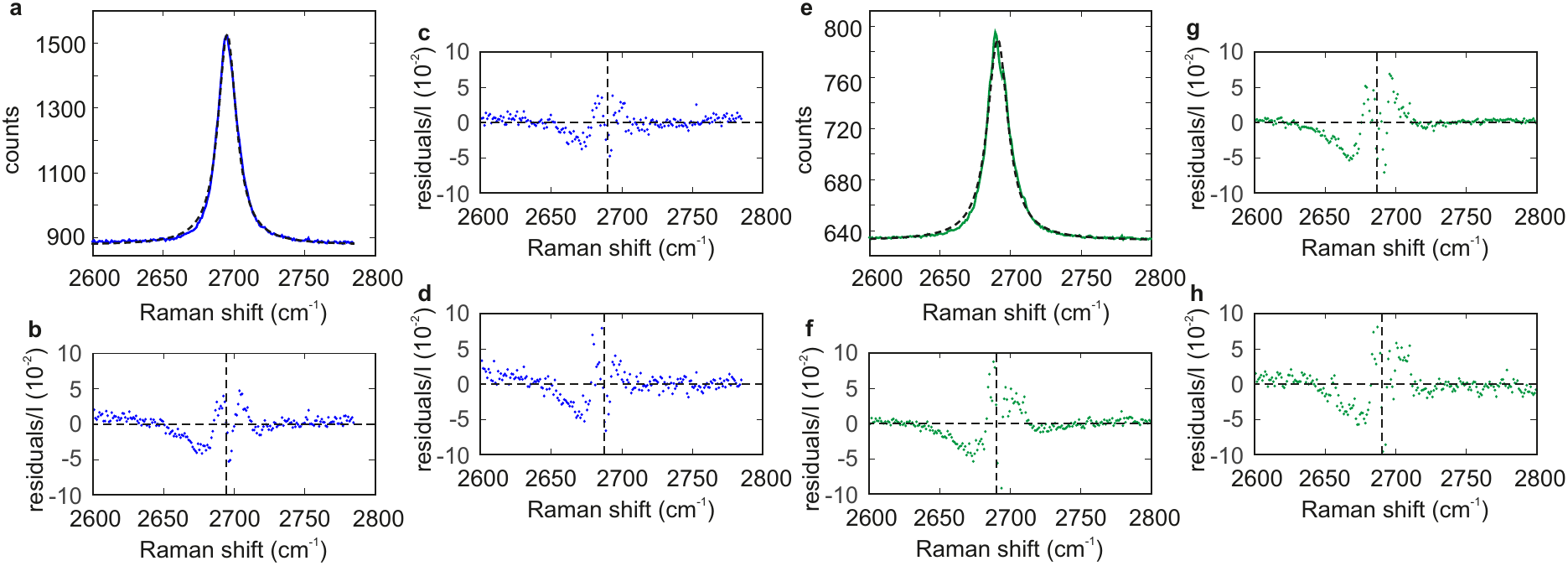}
\caption[Asymmetries in the 2D line shape]{\label{fig:shapeWSe}
(a) Raman spectrum obtained on the hBN-Gr-WSe$_2$ sample (blue) and single Lorentzian fit to the data (black, dashed curve). The fit yields $\Gamma_{2D}=15.8$~cm$^{-1}$. 
(b) Normalized residuals obtained by subtracting the Lorentzian fit from the data of panel (a) and dividing every residual by the peak intensity of the single Lorentzian fit.
(c,d) Residuals of two further spectra measured on the hBN-Gr-WSe$_2$ sample extracted by the method described in (b). The single Lorentzian fit of this spectra yields $\Gamma_{2D}=18.0$~cm$^{-1}$ (c) and $\Gamma_{2D}=16.7$~cm$^{-1}$ (d).
(e) Raman spectrum obtained on the hBN-Gr-hBN sample (green) and single Lorentzian fit to the data (black, dashed curve). The fit yields $\Gamma_{2D}=17.7$~cm$^{-1}$. 
(f) Residuals of the spectrum shown in (d) by the method described in (b).
(g,h) Two further spectra measured on the hBN-Gr-hBN sample extracted by the method described in (b). The single Lorentzian fit of this spectra yields $\Gamma_{2D}=17.0$~cm$^{-1}$ (g) and $\Gamma_{2D}=17.0$~cm$^{-1}$ (h). In all plots of the residuals the horizontal, dashed line marks the zero and the vertical, dashed line indicates $\omega_{2D}$, the center frequencies of the Lorentzian fits. All measurements shown in this figure are based on a laser energy of 2.33~eV.
}
\end{figure*}

\begin{figure}[hbt]
\centering
\includegraphics[draft=false,keepaspectratio=true,clip,width=1\linewidth]{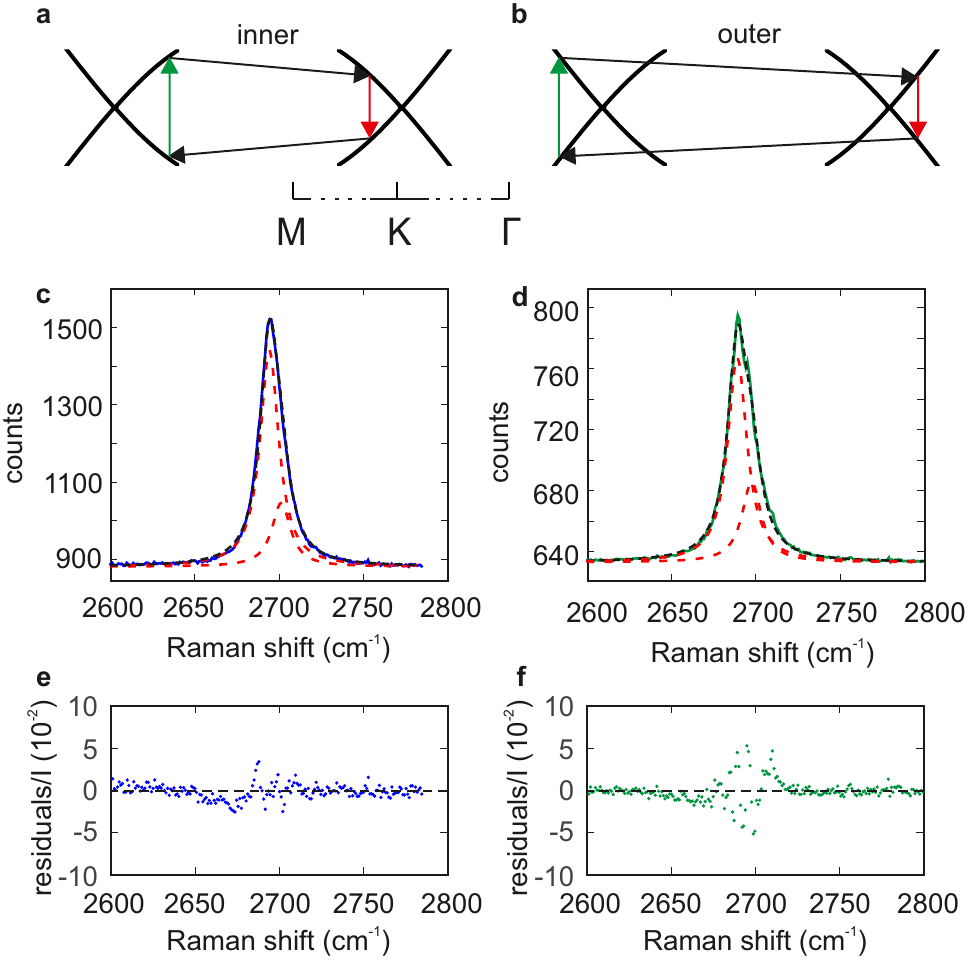}
\caption[Inner and outer processes of the 2D line]{\label{fig:inner}
(a) Schematic of an inner process that leads to one sub-peak of the 2D line. The phonons of the inner process come from the high symmetry line conneting K and $\varGamma$ point in the Brioullin zone.
(b) Schematic of an outer process that leads to one sub-peak of the 2D line. The phonons of the outer process come from the high symmetry line conneting the K and M point in the Brioullin zone.
(c) Double Lorentzian fit to the 2D line obtained on hBN-Gr-WSe$_2$. The measured Raman spectrum is shown in blue. The two sub peaks are shown as red, dashed lines and the combined fit is displayed as a black, dashed curve. The spectrum is the same that is shown in Figure \ref{fig:shapeWSe}a.
(d) Double Lorentzian fit to the 2D line obtained on hBN-Gr-hBN. The measured Raman spectrum is shown in green. The two sub peaks are shown as red, dashed lines and the combined fit is displayed as a black, dashed curve. The spectrum is the same one that is shown in Figure \ref{fig:shapeWSe}e.
(e) Normalized residuals obtained by subtracting the double Lorentzian fit from the data of panel (c) and dividing every residual by the peak intensity of the single Lorentzian fit shown in Figure \ref{fig:shapeWSe}a.
(f) Normalized residuals of the fits in panel (d) obtained in the same way. All measurements shown in this figure are based on a laser energy of 2.33~eV.
}

\end{figure}
Having characterized the doping and strain conditions in both samples, we turn our focus to the line shape of the Raman 2D peak. To study the fundamental properties of the peak, we select spectra with very narrow Raman 2D lines. In Figure \ref{fig:shapeWSe}a a spectrum obtained on the hBN-Gr-WSe$_2$ sample (blue data) and a single Lorentzian fit to the data (black, dashed curve) are shown. Deviations between data and fit are most prominent near the edges of the peak. The asymmetry of the peak is more evident, when plotting the normalized residuals (residuals divided by the peak intensity, $I$,) of the Lorentzian fits which are shown as blue points in Figure \ref{fig:shapeWSe}b (for the residuals, the fit is subtracted from the data). In the measurement, the side at lower wave numbers is steeper than the side at higher wave numbers. Consequently, the subtraction of the symmetric Lorentzian function leads to negative values on the left side and positive values on the right side of the peak in the residual plot. For illustration, the residuals of two further spectra with very narrow 2D line width obtained on the hBN-Gr-WSe$_2$ sample are shown in Figures~\ref{fig:shapeWSe}c and \ref{fig:shapeWSe}d. Again, the asymmetry of the measured 2D lines leads to negative values of the residuals on the left side and positive values on the right side. This behavior is reproducible for many spectra along the sample, but can only systematically be found for very narrow Raman 2D lines (roughly, if $\Gamma_{2D}$ obtained from a single Lorentzian fit is below 20~cm$^{-1}$). This can be well understood by taking into account that the main mechanism for the increased broadening of the Raman 2D line width in these samples are strain variations as analyzed in Reference \citenum{neumann2015b}. The strain distribution within the laser spot is generally unknown and arbitrary, and thus introduces an arbitrary, inhomogeneous broadening to the Raman line. Interestingly, we observe very similar asymmetries in the Raman 2D line of hBN encapsulated, exfoliated graphene. A typical spectrum (green curve) with a single Lorentzian fit (black, dashed curve) is shown in Figure \ref{fig:shapeWSe}e. The corresponding normalized residuals of the fit are presented in Figure \ref{fig:shapeWSe}f, where the asymmetry of the 2D peak becomes again clearly visible. As for the hBN-Gr-WSe$_2$ sample, this behavior is reproducible over the entire sample with two further residual plots obtained on different positions on the graphene layer shown in Figures \ref{fig:shapeWSe}g and \ref{fig:shapeWSe}h.

\begin{table*}[hbt]
\begin{centering}
\begin{tabular}{ccccc}
\hline
    \hline
    substrate & $\omega_{2D+}-\omega_{2D-}$~(cm$^{-1}$)  & $I_{2D-}$/I$_{2D+}$ & $\Gamma_{2D,\pm}$~(cm$^{-1}$) \\
    \hline
    hBN-Gr-WSe$_2$ & $6.6 \pm 0.5$ & $2.5 \pm 0.8$ & $14.2 \pm 1.0$ \\
    hBN-Gr-hBN & $8.9 \pm 1.0$ & $3.9 \pm 2.1$ & $14.4 \pm 1.0$ \\
		suspended \cite{berciaud2013} & $13.9 \pm 0.5$  & $3.2$ & $17.5 \pm 0.5$ \\
    \hline
\end{tabular}
\par\end{centering}
\centering{}
	  \caption{Average values from the double Lorentzian fits for the two different samples are compared to data obtained by Berciaud \textit{et al}. \cite{berciaud2013} on suspended graphene.}
  \label{tab:lineshape}
\end{table*}

So far, clear asymmetries in the line shape of pristine graphene have only been reported for suspended graphene \cite{berciaud2008,luo2012,berciaud2013}. In these studies the asymmetry was attributed to a co-existence of the inner- and outer Raman processes (Figures \ref{fig:inner}a and \ref{fig:inner}b) involving phonons of different energies due to trigonal warping as proposed by theoretical calculations \cite{venezuela2011}. As a consequence, the inner and outer processes are believed to lead to two distinguishable sub-peaks that add up to the measured Raman 2D peak \cite{luo2012}. More specifically, the inner processes originate from phonons that come from the region between the K and $\varGamma$ point of the Brioullin zone, while the outer processes involve phonons from the region between the K and M point.
In Figures \ref{fig:inner}c and \ref{fig:inner}d we present double Lorentzian fits to the 2D peaks taken on our hBN-Gr-WSe$_{2}$ and hBN-Gr-hBN samples (same spectra as in Figures \ref{fig:shapeWSe}a and \ref{fig:shapeWSe}e, where we used the frequencies $\omega_{2D+}$ and $\omega_{2D-}$, the intensities of the sub-peaks $I_{2D+}$ and $I_{2D-}$ and a common broadening $\Gamma_{2D,\pm}$ for both sub-peaks and an offset as fit parameters. 
The results of our double Lorentzian fits are summarized in Table \ref{tab:lineshape}. In particular,  we report the averages of the intensity ratio, $I_{2D-}/I_{2D+}$, and frequency difference, $\omega_{2D+}-\omega_{2D-}$, obtained from the three spectra on each material stacks that are analyzed in Figure \ref{fig:shapeWSe}. The intensity ratio varies strongly from spectrum to spectrum, making concise statements rather difficult. However, the averages seem to coincide with the values reported in Reference \citenum{berciaud2013}. In contrast, the frequency difference can be obtained more reliably and with far smaller variation between individual spectra from our data. We find $\omega_{2D+}-\omega_{2D-} = 6.6 \pm 0.5$~cm$^{-1}$ for hBN-Gr-WSe$_2$ and $\omega_{2D+}-\omega_{2D-} = 8.9 \pm 1.0$~cm$^{-1}$ for hBN-Gr-hBN. The values are similar for both substrate combinations, but remarkably lower than the findings made by Berciaud \textit{et al}. \cite{berciaud2013} on suspended graphene at the same laser wavelength, who obtained a sub-peak splitting of around 14~cm$^{-1}$. Theoretically, Venezuela \textit{et al}. expect a mostly vanishing splitting of the sub-peaks due to a cancellation of trigonal warping effects of the phonon and electronic dispersion relation at 2.33~eV from first principle calculations \cite{venezuela2011}. This would result in a symmetric 2D line shape, which is not in agreement with our findings. In general, the description with two Lorentzians captures the data better than the single Lorenzian fits for all investigated spectra. Still, we find small, systematic deviations between the data and double Lorentzian fits in many spectra as evident from Figures \ref{fig:inner}e and \ref{fig:inner}f, where we show the normalized residuals of the fits shown in Figures \ref{fig:inner}c and \ref{fig:inner}d. This could indicate that contributions solely from inner and outer process from lines connecting the high symmetry points in the Brioullin zone are not sufficient to accurately describe the Raman 2D peak, supporting the deductions made in Reference \citenum{berciaud2013}, where Berciaud \textit{et al}. concluded from a laser dependent analysis of suspended graphene that fully two dimensional calculations taking into account the entire Brioullin zone around the K and K' valleys might be required to accurately describe the Raman 2D line.

\section{Conclusion}
In summary, we investigated the line shape of the Raman 2D line of exfoliated and CVD graphene in van der Waals heterostructures, where hBN and WSe$_2$ were used as substrate materials. In both cases, we observed extremely low doping of the graphene crystal and high strain uniformity on sub-laser spot length scales, which is reflected in the very narrow minimal line width of the 2D lines down to 15-16~cm$^{-1}$. Notably, these values are significantly smaller than the lowest values for the line width obtained on suspended graphene samples (that the authors are aware of). For both van der Waals heterostructures, the spectra with the narrowest Raman 2D peaks showed clear asymmetries in this line shape. We fitted double Lorentzian functions to the spectra to account for the inner and outer processes that are expected to contribute to the measured Raman 2D line. For the two sub-peaks, we find an average splitting of $6.6 \pm 0.5$~cm$^{-1}$ (hBN-Gr-WSe$_2$) and $8.9 \pm 1.0$~cm$^{-1}$ (hBN-Gr-hBN), which is smaller as compared to the splitting measured on suspended graphene at the same laser energy. Although the double Lorentzian functions describe the measured peaks better than the single Lorentzian functions, they still show small systematic deviations. This could indicate that contributions from parts of the Brioullin zone away from the high symmetry axes also contribute to the Raman 2D peak and that fully two dimensional calculations are required to accurately describe the Raman process.

\section{Acknowledgements}
Support by the Helmholtz Nanoelectronic Facility (HNF), the Deutsche Forschungsgemeinschaft, the ERC (GA-Nr. 280140), and the EU project Graphene Flagship (contract no. NECT-ICT-604391), are gratefully acknowledged.
S. Reichardt acknowledges funding by the National Research Fund (FNR) Luxembourg. K. Watanabe and T. Taniguchi acknowledge support from the Elemental Strategy Initiative conducted by the MEXT, Japan and a Grant-in-Aid for Scientific Research on Innovative Areas “Science of Atomic Layers” from JSPS.

\providecommand{\WileyBibTextsc}{}
\let\textsc\WileyBibTextsc
\providecommand{\othercit}{}
\providecommand{\jr}[1]{#1}
\providecommand{\etal}{~et~al.}

\end{document}